\documentclass[aps,twocolumn,toolkits,nofootinbib]{revtex4}
\usepackage{mathrsfs}
\usepackage{amsmath}
\usepackage{graphicx}
\usepackage{color}
\usepackage{epsfig}
\usepackage{subfigure}

\begin{document}

\title{The Schwinger mechanism and graphene}

\author{Danielle Allor}
\email{dallor@umd.edu}

\affiliation{Department of Physics, University of Maryland,
College Park, MD 20742-4111}

\author{Thomas D. Cohen}
\email{cohen@physics.umd.edu}

\affiliation{Department of Physics, University of Maryland,
College Park, MD 20742-4111}

\author{David A. McGady}
\email{dmcgady@princeton.edu}

\affiliation{Department of Physics, Princeton University, , Princeton, NJ 08544}


\begin{abstract}
The Schwinger mechanism, the production of charged particle-antiparticle pairs in a macroscopic external electric field, is derived for 2+1 dimensional theories.  The rate of pair production per unit area for four species of massless fermions, with charge $q$, in a constant electric field $E$ is given by $ \pi^{-2} \, \hbar^{-3/2} \,\tilde{c}^{-1/2}\, ( q E)^{3/2} $ where $\tilde{c}$ is the speed of light for the two-dimensional system.  To the extent undoped graphene behaves like the quantum field-theoretic vacuum for massless fermions in 2+1 dimensions, the Schwinger mechanism should be testable experimentally.  A possible experimental configuration for this is proposed.  Effects due to deviations from this idealized picture of graphene are briefly considered. It is argued that with present day samples of graphene, tests of the Schwinger formula may be possible.
\end{abstract}

\maketitle

\section{Introduction}

The Schwinger mechanism refers to the production of charged particle-antiparticle pairs out of the vacuum by a classical electric field which is homogenous over a large volume in space. More than half a century ago, Schwinger computed\cite{Schwinger} the rate associated with vacuum breakdown via pair production per unit volume for the case of a constant electric field. Over the years, the Schwinger mechanism has spawned a vast literature. Invoked to gain insights on topics as diverse as the string breaking rate in QCD\cite{CNN,Neuberger} and on black hole physics\cite{GR}, this mechanism has become a textbook topic in quantum field theory\cite{IZ}.  Topics such as back reaction\cite{BR} and finite size effects\cite{Wang_Wong} have been addressed.

A key issue is the rate of pair production from a static electric field.  Schwinger originally addressed this question by noting that the probability that a system with zero electrons or positrons remains in the fermionic vacuum state decays exponentially in time\cite{Schwinger}. Treating the electric field classically---\emph{i.e.} the formal limit of $q\rightarrow 0$, $E \rightarrow \infty$ with $q E$ fixed---Schwinger  calculated the vacuum persistence probability, $P_{\rm vac}(t)$, as a function of time:
\begin{eqnarray}
P_{\rm vac}(t)& \equiv& |\langle {\rm vac}| U(t) |{\rm vac} \rangle |^2 = \exp(-w V t) \label{SF1}\\
{\rm with} \; \; w &=& \frac{(q E)^2}{4 \pi^3 \hbar^2 c} \, \sum_{n=1}^\infty  \frac {1}{n^2} \, \exp \left( -\frac{n \pi m^2 c^3}{q E \hbar}\right ), \label{SF2}
\end{eqnarray}
where $V$ is the spatial volume of the system and $w$ is the rate of vacuum decay per unit volume.

Schwinger's interpretation\cite{Schwinger} of Eqs.~(\ref{SF1}) and (\ref{SF2}) was straightforward: $w$ was taken to be the local rate of production per unit volume of fermion-antifermion pairs  by the electric field. This interpretation has been widely accepted in much of the literature on the Schwinger mechanism.  However, while the Schwinger formula of Eq.~(\ref{SF2}) is very well known, it has been argued that its interpretation as the pair production rate is not correct\cite{Niki}.  Despite the very natural interpretation of $w$ in Eq.~(\ref{SF2}) as the rate of production of pairs per unit volume, an explicit calculation gives the rate of pair production per unit volume, which we denote $\Gamma$, as
\begin{equation}
\Gamma = \frac{(q E)^2}{4 \pi^3 \hbar^2 c} \, \exp \left( -\frac{\pi m^2 c^3}{q E \hbar} \right ) \; . \label{SF3}
\end{equation}
This rate does {\it not} agree with $w$: the entire rate is given by the first term in the series for $w$. For a recent, pedagogical, discussion highlighting the theoretical distinction between these two rates, see Ref. \cite{Cohen_McGady}.

Despite its theoretical significance, there has been no direct experimental signature of the Schwinger mechanism, of charged pair creation in electric fields.  This is particularly unfortunate given the common confusion between the rate of pair creation $\Gamma$ and $w$ the rate of vacuum decay.   It would be {\it very} useful to concoct an experimental test to distinguish between the two directly.  Moreover, apart from the distinction between $w$ and $\Gamma$ the derivation of the rate of pair production via the Schwinger mechanism raises a number of subtle issues associated with the implementation of appropriate boundary conditions\cite{GR,GG,Wang_Wong,Neuberger}; it is important to test whether these are handled correctly.  Ultimately the most compelling test would be experimental.

The reason that the Schwinger mechanism has never been tested experimentally is very easy to understand: the exponential factor in the pair production rate is {\it very} small for static macroscopic $E$ fields realizable in the lab. It only becomes of order unity when $E$ is large enough so that $q E$ times the electron's Compton wavelength is greater than $m c^2$: this requires an electric field of order $10^{16} {\rm V/cm}$; for $E = 10^6 \frac{\rm V}{\rm cm}$, $\exp \left(-\frac{ \pi m^2 c^3}{q E \hbar} \right ) \approx \exp\left(-4 \times 10^{10} \right)$.

One might hope to test the Schwinger formula experimentally in a condensed matter system which simulates light or massless, electrically charged, relativistic fermions. In this context, the condensed matter system acts as an analog computer to test the underlying result from relativistic field theory. Fortunately, it has been known for more than two decades that charged quasi-particle excitations in a potential with a two-dimensional hexagonal symmetry have a region of momenta over which their dispersion relation is linear-$(\epsilon - \epsilon_0)^2 = \tilde{c}^2 (p_x^2 + p_y^2)$\cite{Semenov}. This is precisely the dispersion relation of a massless relativistic particle with energy measured relative to $\epsilon_0$ and has $\tilde{c}$ playing the role of $c$. Graphene ({\it i.e.}, a single sheet of graphite) has such a symmetry. Moreover, in undoped graphene, the Fermi level is at $\epsilon_0$. Thus, to the extent that a single particle description holds in graphene, the quantum ground state of a filled Fermi sea is the precise analog of a filled Dirac sea---{\it i.e.} the vacuum of a two-dimensional non-interacting field theory for fermions. The recent development of techniques to produce samples of graphene and measure its properties\cite{GrExp,GrExp2} has focused significant attention to its analogy with massless Dirac particles: graphene has been proposed as a testing ground for the standard relativistic quantum mechanical effects of zwitterbewegung and Klein/Landau-Zener tunneling \cite{KP,GPNJ}. This letter explores the possibility of using graphene to test experimentally the more subtle dynamics of the Schwinger mechanism.

\section{The Schwinger pair creation rate in graphene}

One can adapt Schwinger's calculation\cite{Schwinger} for smaller dimensions\cite{1plus1, 2plus1,GG}; in (2+1) dimensions the probability that the system has remained in the (fermionic) vacuum after time $t$, $P_{\rm vac} (t)$, is
\begin{eqnarray}
P_{\rm vac}^{2+1}(t)& =&  \exp(-w^{2+1} A t) \\
{\rm with} \; \; w^{2+1} &=& \frac{f \, (q E)^{3/2}}{4 \pi^2 \, \hbar^{3/2} \, \tilde{c}^{1/2}} \, \sum_{n=1}^\infty  \frac {1}{n^{3/2}}\,
\exp \left( -\frac{n \pi m^2 \, \tilde{c}^3}{q E \, \hbar} \right )\nonumber \\
& =&  \frac{f \, \zeta \left ( \frac{3}{2} \right )\,( q E)^{3/2}}{4 \pi^2 \, \hbar^{3/2} \, \tilde{c}^{1/2} } \; \; \; \;  {\rm for}  \; \;  m=0 \label{2plus1w}
\end{eqnarray}
where $A$ is the spatial area, $\tilde{c}$ is the speed of light for the 2-d system, $\zeta$ is the Riemann zeta function with $\zeta (3/2) \approx 2.612$, and $f$ is the number of species of fermion ({\it i.e.}, four for graphene). Similarly, the local rate of pair creation, $\Gamma^{2+1}$, is given by the first term of this series\cite{Cohen_McGady,Niki,GG}, 
\begin{equation}
\Gamma^{2+1} = \frac{f (q E)^{3/2}}{4 \pi^2 \, \hbar^{3/2} \, \tilde{c}^{1/2}} \exp \left( -\frac{n \pi m^2 \, \tilde{c}^3}{q E \, \hbar} \right ).  \label{2plus1G}
\end{equation}
These derivations hold for classical, constant, and externally fixed electric fields. Consequently these relations are valid only when it is legitimate to neglect: i) macroscopic back reaction to the applied field due to this same charged particle production rate\cite{BR}; ii) the production of real photons as the charged particles accelerate in the electric field; and iii) interactions between the fermions mediated by the exchange of virtual photons.

Alternative derivations\cite{CNN,Cohen_McGady} to Schwinger's original calculation\cite{Schwinger}, focus on the nature of a single particle level for the Dirac equation in the presence of an electric field, switched on in the distant past. These derivations have the twin virtues of clarifying both the role of boundary conditions, and the distinction between the rate associated with vacuum breakdown, $w$, and the rate of pair creation, $\Gamma$. Here we will discuss the derivation done in the time independent gauge, $E_i = -\partial_i A_0(x_j)$\cite{CNN,Wang_Wong}. In this gauge, we see that the electric field alters the single particle energy levels. The shifts in energy due to the potential have opposite signs on either side of the field region, allowing filled levels on one side to become degenerate with empty ones on the other. This yields an effective potential which filled levels in the elevated Dirac sea tunnel through, leaving a hole on one side and yielding a particle type state on the other.

This tunneling  problem is conceptually simple for an electric field of limited spatial extent\cite{Wang_Wong}. Consider an infinite plane with an electric field independent of $y$, with magnitude $E$, oriented in the $-x$ direction and confined to the region between $-L/2 \le x\le L/2$. A useful basis is the in-state wave functions which correspond to solutions of the Dirac equation with unit flux moving towards the region of the electric field from either the left or right. The states have  amplitude $T$ (times a flux-normalizing kinematic factor) to be found on the far side of the field region.  $T$ may be computed directly as a tunneling problem in an effective energy-dependent Schr\"odinger equation derivable from the underlying Dirac equation.

\begin{figure}
    \centering
     \includegraphics[width=3in]{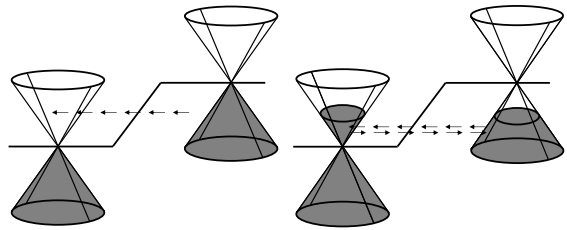}
      \caption{The dispersion relations are shifted on either side of the electric field. The image on the left is a cartoon picture of the in-state vacuum, where the Schwinger mechanism applies. After $\tau_{fill}$ the level occupations change, and the system falls into a p-n type system, depicted on the right.}
       \label{electronic_structure}
\end{figure}

Suppose that the system is in the ``in-state vacuum'' (the left image in Fig.~\ref{electronic_structure}).  All of the the in-states below the Dirac sea on the left, $\psi^{L \, \rm in}_{\epsilon,p_T}$, are occupied for $\epsilon < - q E L/2$ and all of the states below the Dirac sea on the right, $\psi^{R \, \rm in}_{\epsilon,p_T}$, are occupied for $\epsilon < q E L/2$. Turning on the electric field in the distant past has merely shifted the energies of the states relative to the local vacuum, not their occupation number.  The pair production rate for a particle with energy $\epsilon$ and transverse momentum $p_T$ is proportional to the transmission probability for a filled in-state from the left $-q E L/2 \le \epsilon \le q E L/2$. In essence a filled level moving in the elevated Dirac sea towards the region of the electric field propagates through and emerges on the other side, with probability $|T|^2$, where it appears as a particle. The rate of pair production per unit width per unit $\epsilon$ per unit $p_T$ is computed analogously to the 3+1 dimensional case\cite{CNN}. Integration over $\epsilon$ and $p_T$ gives the rate of pairs per unit width:
\begin{eqnarray}~
\frac{d^2 N}{ d t\, d W } & = & \int_{-\epsilon^{\rm max}}^{\epsilon^{\rm max}} d \epsilon \int_{-p_T^{\rm max}}^{p_T^{\rm max}} d p_T \frac{d^4 N}{d \epsilon \, d p_T \, d W \, d t} \nonumber \\
\epsilon^{\rm max} =   \frac{q E L}{2}&-&m \tilde{c}^2 \; \; , \; \; p_T^{\rm max} = \frac {\sqrt{(2 \epsilon -q E L)^2 - 4 m^2 \tilde{c}^4}}{2 \tilde{c}} \nonumber \\
\frac{d^4 N}{d \epsilon \, d p_T \, d W \, d t} & = &  f \frac{|T(\epsilon,p_T)|^2 }{4 \pi^2 \, \hbar^2}   \; .\label{rate}
\end{eqnarray}
In the large $L$ limit, the WKB result is increasingly valid, and $|T|^2= \exp (-\frac{\pi (m^2+p_T^2) \, \tilde{c}^3}{q E \, \hbar})$ \cite{CNN}. Evaluating $T$ within the WKB approximation for arbitrary $L$, and equating $ \frac{d^2 N}{ d t\, d W } $ with $\Gamma_{2+1}$, one arrives at the 2+1 dimensional Schwinger formula of Eq.~(\ref{2plus1G}) for $L^2 \gg \hbar \tilde{c}/q E .$ If $L$ is not large one must compute $T$ from the Dirac equation and numerically integrate over $\epsilon$ and $p_T$. The ratio of the rate at finite $L$ to the Schwinger rate depends only on the ratio $L / \sqrt{\hbar \tilde{c} /q E }$. The numerically calculated rate smoothly converges to the Schwinger rate; see Fig.~\ref{finiteL}.

It should be noted that while the preceding derivation is modeled on the derivation (in 3+1 dimensions) of Ref. \cite{Wang_Wong} there is one crucial distinction: Ref.~\cite{Wang_Wong} computed $\Gamma$ (the vaccum decay rate) which was implicitly assumed to be the pair production rate $w$.  This difference only appears in the last line of Eq.~(\ref{rate}): to compute $w_{2+1}$ rather than $\Gamma_{2+1}$ one replaces $-\log ( 1- |T(\epsilon,p_T)|^2 )$ (used in Ref. \cite{Wang_Wong}) with $|T(\epsilon,p_T)|^2$.

\section{A proposed experiment}

The previous derivation depends on the system  being in the in-state vacuum. Intuitively, after a certain transient time during which the system equilibrates, the rate should be dominated by incoming levels from far away.  Since the time scale for such transient behavior is finite at infinite $L$\cite{Neuberger}, the natural time scale associated with transients does not depend on $L$.   This transient time scale can be obtained via simple dimensional analysis: $\tau_{\rm trans} = \sqrt{{\hbar}/{(\tilde{c} \, q E)}}$.

\begin{figure}
   \centering
       \includegraphics[width=3in]{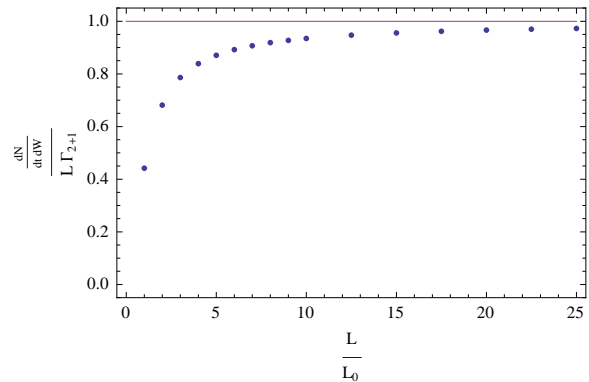}
       \caption{The ratio of the pair production rate at finite $L$ to the (infinite $L$) Schwinger formula rate for a 2+1 dimensional system. $L$ is measured in units of $L_0 \equiv {\sqrt {\hbar \tilde{c} / (q E)}}$. }
       \label{finiteL}
\end{figure}

The analysis crucially requires the fermionic ``vacuum'' to be placed in an external electric field which remains constant over time. To experimentally realize this with graphene, one must fix an external voltage by placing a region of the graphene sheet between two conducting plates, held at {\it constant} voltage. Pair production is driven by differences in level occupation on either side of the field. Particles and holes created in the field region are carried into the regions of the graphene sample on either side of the field regions which serve as {\it reservoirs} for particles and holes. After a finite time, the accessible states fill substantially and the system leaves the regime of validity of the Schwinger formula; the reservoirs develop an excess of holes or particles and the system more closely resembles p-n junctions, under extensive study in the context of Landau-Zener tunneling\cite{GPNJ}; see Fig.~\ref{electronic_structure}.  The natural timescale needed to deplete a substantial fraction of the reservoir and depart from the Schwinger regime is $\tau_{\rm fill} \equiv \sqrt{ q E/(\hbar \tilde{c}^3)}\, L L_{\rm res}$.

\begin{figure}
   \centering
      \includegraphics[width=3.3in]{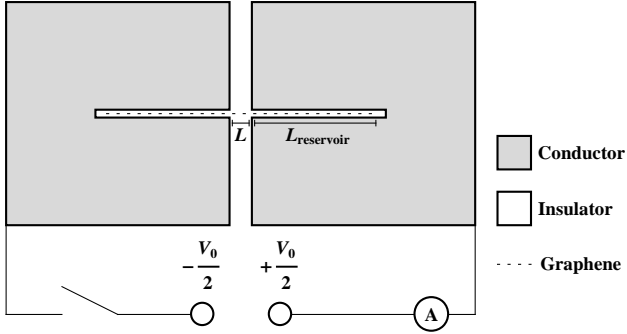}
       \caption{Schematic depiction of the cross-sectional view of a possible experiment measuring the rate of production of Schwinger pairs.    }
       \label{experiment}
\end{figure}

An experiment to test the Schwinger formula is thus conceptually straightforward: a sheet of graphene of width $W$, and length $2 L_{\rm res} + L$ is placed in an apparatus whose cross-section is given schematically in Fig.~\ref{experiment}; $L$ is the length the field region and $L_{\rm res}$ is the length of the reservoirs to either side. An electric field of magnitude $V_0/L$  is turned at $t=0$; the Schwinger formula (\ref{2plus1G}) should be accurate for $\tau_{\rm trans } \ll t \ll\tau_{\rm fill}$, bringing about a two-dimensional current density ${\cal J}$ from the Schwinger pairs. For $\tau_{\rm trans } \ll t \ll \tau_{\rm fill}$, the current density just beyond the field regions to good approximation is
\begin{equation}
{\cal J}_{\rm Sch}\equiv q \, \Gamma^{2+1} \ L =  \, \frac{ q \,( q V_0)^{3/2}}{ \pi^2 \, \hbar^{3/2} \, \tilde{c}^{1/2} \, L^{1/2} } \; . \label{dens}
\end{equation}
As charge flows into the reservoirs, an external current $I={\cal J}_{\rm Sch} W$ must flow to the conductors to maintain them at fixed voltages of $\pm V_0/2$. This current $I$ can be monitored to determine ${\cal J}$. Note that the graphene sheet is fully insulated electrically and is {\it not} part of a closed circuit. Thus, the current is necessarily transient and the experiment does {\it not} measure the usual conductance. This proposed method of testing Schwinger mechanism in graphene has a key feature in common with conductance in graphene p-n junctions (GPNJ) in the ballistic regime: both systems depend on quantum tunnelling of massless Dirac particles\cite{GPNJ}.

Effects due to non-zero temperature, non-zero lattice spacing, impurities, finite size effects and temporal transients present in real systems can affect the results of the measurements. One expects to be in the regime of validity for the Schwinger formula if the parameters satisfy:
\begin{eqnarray}
\sqrt{\frac{\, q E_0 }{\hbar \tilde{c}^3}} \, L \,  L_{\rm res}&\gg& t \gg  \sqrt{\frac{\hbar  }{q E_0\tilde{c}}} \label{tcond}\\
L , \,  W , \,  l_{\rm mfp} & \gg &\sqrt{\frac{\hbar \tilde{c}}{q E_0}} \label{Lcond}\\
V_0 & \ll & \frac{\hbar \tilde{c} }{ q a} \approx 2.5 V  \label{Vcond}\\
T & \ll & \sqrt[4]{\hbar \tilde{c} ( q E_0)^3 L^2/k_B^4} \; ,  \label{Tcond}
\end{eqnarray}
where $E_0 \equiv V_0/L$, $a$ is the lattice spacing and $l_{\rm mfp}$ is an effective mean-free path. Conditions (\ref{tcond}) and (\ref{Lcond}) for $L$ relate to time transient effects and finite length issues associated with the WKB formalism, respectively; they are discussed above. The analogous condition for $W$ follows from the requirement that the discrete mode sum in $p_T$ is approximated by the Gaussian integral in Eq.~(\ref{rate}). Since the distance scale for the creation of pairs in the Schwinger mechanism is $L_0 \equiv \sqrt{\hbar \tilde{c}/ (q E)}$, the Schwinger formula only applies when the dynamics are well described by the simple Dirac description over that scale, requiring $l_{mfp} \gg L_0$. Condition (\ref{Vcond}) encodes the requirement of a linear dispersion relation; this fails for momenta comparable to $a$. Finally, Condition (\ref{Tcond}) applies to thermal fluctuations, which imply a density of particles and holes in the reservoirs.  These can randomly wander into, and get transported across, the field region, creating a current. The condition ensures that the Schwinger current dominates.

The experimental configuration does not directly measure conductivity since current flow is necessarily transient. Still one might worry that the dynamics associated with the usual conductivity could mask the Schwinger effect. However, Eqs. (\ref{2plus1G}) and (\ref{dens}) imply that ${\cal J}_{\rm Sch} = \left( \frac{4 q^2 E_0}{h} \right ) \left [ \frac{ 1}{2 \pi}\right ] \sqrt{ \frac{L^2 q E_0}{\hbar \tilde{c}}}$.  The term in parenthesis is of the scale expected from standard conductivity mechanisms: in graphene it is $\sigma \sim 4 q^2/h$\cite{GrExp2}. The term in square brackets is of order unity; Condition (\ref{Lcond}) implies that the square root factor is large. Thus, the transient current density induced by the Schwinger mechanism dominates over what is expected from the usual conductivity.

Whether Conditions (\ref{tcond})-(\ref{Tcond}) can be satisfied in practice depends critically on both the size of the graphene sample and its purity.  In practice, with samples with sizes restricted to $\sim 100 \, \mu {\rm m}$ and impurity concentrations reported in \cite{GrExpMFP, rev}, the conditions appear to require extremely fast measurements but do not seem to be beyond current technology. For the purpose of making estimates we will assume that the size of the sample is $\sim 100 \, \mu {\rm m}$; for concreteness we will take $W =100 \, \mu {\rm m}$, $L =1 \, \mu {\rm m}$ and $L_{\rm res}= 49 \, \mu {\rm m}$. We also take $V_0 =1$V which effectively satisfies Condition (\ref{Vcond}) \cite{rev}. With the values above,$\sqrt{\hbar \tilde{c}/q E_0} = 25 {\rm nm}$, and Condition (\ref{Lcond}) is satisfied by a factor of 40 for $L$ and a factor of 4000 for $W$.  Condition (\ref{Tcond}) is also well satisfied provided $T \ll 1800^\circ$K.

Condition (\ref{Lcond}) for $l_{mfp}$ depends on a quasi-particle's energy, $\epsilon$.  It has been argued with current samples the dominant contribution is from Coulomb impurities\cite{GrExpMFP}. For the purposes of estimating $l_{mfp}$, this will be assumed to be correct. Coulomb scattering has an infinite cross-section and the mean-free path is not well defined. However, an effective mean free path in the sense of the characteristic distance a quasiparticle travels before it is substantially affected by the impurities can be estimated. It is the distance a quasi-particle travels before entering a region in which the Coulomb energy is comparable to the kinetic energy, yielding $l_{mfp} \sim \kappa \epsilon (\hbar \, c \, \alpha  \, n_{imp})^{-1}$ where $\kappa$ the dielectric constant of the insulator and $n_{imp}$ is the density of impurities.  For Schwinger pairs, $\epsilon$ is of order (but less than) $q V_0$; Condition (\ref{Lcond}) becomes $ \frac{ \kappa \epsilon}{\hbar c \alpha n_{imp}}  \gg \sqrt{\frac{\hbar \tilde{c} L}{q V_0}}$. Samples with $n_{imp}$ as small as $\sim 2 \times 10^{11} cm^{-2}$ have been reported\cite{{GrExpMFP}}. Using this value, the left-hand side of the inequality is 350 $\kappa$ nm. The right-hand side is 25 nm. Thus, the condition appears to be satisfied at least moderately well.  Using the results in Fig.~\ref{finiteL} as a guide in estimating errors suggests that at the least, a semi-quantitative test of the Schwinger mechanism should be possible. If a substrate with relatively large $\kappa$ proves viable, the condition may be well-satisfied. It appears possible that the mean-free paths are long enough to allow for a meaningful test of the Schwinger mechanism with currently available samples.  Precision tests will probably require cleaner samples which one hopes may become available in the future.

With the parameters given above, $\tau_{fill} \equiv \sqrt{\frac{\, q V_0 L}{\hbar \tilde{c}^3}}  \, L_{\rm res} \approx 1.9 \times 10^{-9} {\rm s}$ while $\tau_{trans}\equiv \sqrt{\frac{\hbar L }{q V_0\tilde{c}}} \approx 2.6 \times 10^{-14} {\rm s}$.  It is easy to ensure $t \gg \tau_{trans}$. The restriction $t \ll \tau_{fill}$, however, requires taking data at a very high rate---considerably faster than one GHz. Fortunately, it {\it is} possible to take data at rates much faster than one GHz. If future sample sizes increase significantly, one could increase the size of $L_{res}$ and thereby $\tau_{fill}$ and thus reduce the technical challenges associated with very rapid measurements.

\section{Conclusion}

To summarize, even with presently available samples, there is good reason to believe that the regime of validity of the Schwinger formula can be realized experimentally, at least at a semi-quantitative level.  As larger and higher quality samples become available in the future, practical tests of the Schwinger formula with increasing accuracy ought to become possible.  It is reasonable to expect that such experimental probes should become sufficient to test quantitatively the pair production of the Schwinger mechanism; it is is important that such measurements are accurate enough to distinguish between the predicted rate of pair of production and the rate of vacuum decay.

{\it Acknowledgments.} The authors benefitted greatly from discussions with S.~Adam, S.~Das Sarma, M.~Fogler, M.~Furher, V.~Galitsky, S.P.~Gavrilov, T.~Jacobson, K.G.~Klimenko, K.S.~Novoselov and I.A.~Shovkovy. D.A. and D.A.M were supported by the University of Maryland through its Senior Summer Scholars program. T.D.C.\ was supported by the United States D.O.E.\ through grant number DE-FGO2-93ER-40762.


\begin{thebibliography}{99}

\bibitem{Schwinger}  J.~Schwinger,  Phys.\ Rev.\  {\bf 82}, 664 (1951).

\bibitem{IZ}   C.~Itzykson and J.-B.~Zuber, {\it Quantum Field Theory}, McGraw-Hill (1980).

\bibitem{CNN} A. Casher, H Neuberger and S. Nussinov, Phys.\ Rev.\ D\ {\bf  20}, 179 (1979).

\bibitem{Neuberger} H. Neuberger , Phys.\ Rev.\ D\ {\bf  20} 2936 (1979).

\bibitem{BR} Much of the theoretical interest over the years has been on the encorportation of backreaction.  See for example, F.~Cooper and E.~Mottola, Phys.\ Rev.\ D\ {\bf 40}, 456 (1989); Y.~ Kluger,J.M.~Eisenberg, B.~Svetitsky,F.~Cooper and E.~Mottola, Phys.\ Rev.\ lett\ {\bf 67}, 2427 (1991); Phys.\ Rev.\ D\ {\bf 45}, 4659 (1992) and ref.~\cite{GR}.

\bibitem{Wang_Wong}   R.C.~Wang and C.Y.~Wong,  Phys.\ Rev.\ D\ {\bf 38}, 348 (1988).

\bibitem{2plus1}K.G.~Klimenko, Z.~Phys.~ {\bf C 54}, 323 (1992); K.G.~Klimenko, Theor.~Math.~Phys. {\bf 89}, 1287 (1992); W.~Dittrich and  H.~Gies, Phys.~Lett.~ {\bf B 392}, (1997), 182; V.P.~Gusynin and I.A.~Shovkovy, J.~Math.~Phys. {\bf 40}, 5406 (1999); S. Maroz, arXiv:0710.4880v1[hep-ph].

\bibitem{1plus1} Q.-g. Lin, J. Phys. G. Part. Phys. {\bf 25}, 17 (1999).

\bibitem{GR} R.~Brout, S.~Massar, R.~Parentani, S.~Popescu and Ph.~Spindel, Phys.\ Rev.\ D\ {\bf 52}, 1119 (1994).

\bibitem{GG}  S.P.~Gavrilov and D.M.~Gitman, Phys.\ Rev.\ D\ {\bf 53}, 7162 (1996).

\bibitem{Cohen_McGady} T.D.~Cohen, D.A.~McGady, Phys.\ Rev.\ D\ {\bf 78}, 036008 (2008).

\bibitem{Niki} A.I.~Nikshov, Nucl.\ Phys.\ {\bf B 21}, 346 (1970).


\bibitem{Semenov} G.W.~Semenov, Phys.\ Rev.\ lett\ {\bf 53}, 2449 (1984).

\bibitem{GrExp} K.S.~Novoselov {\it et al}, Science {\bf 306}, 666 (2004); K.S.~Novoselov {\it et al}, Proc.\ Natl.\ Acad.\ Sci.\ {\bf 102}, 10451 (2005);  Y.~Zhang, Y.-W.~Tan, H.L.~Stormer and P.~Kim, Nature {\bf 438}, 197 (2005).

\bibitem{GrExp2}K.S.~Novoselov {\it et al}, Nature {\bf 438}, 197 (2005).

\bibitem{KP}  M.I.~Katsnelson, Eur.\ Phys.\ J.\  {\bf B 51}, 157 (2006); M.I.~Katsnelson, K.S.~Novoselov and A.K. Giem Nature Phys.\ {\bf 2}, 620 (2006); J.M.~Perira, V.~Mlinar, F.M. Peeters, and P.~Vasiloulos,  Phys.\ Rev.\ B\ {\bf 74}, 045424 (2006); M.I.~Katsnelson and K.S. Novoselov, cond-mat/0703374.

\bibitem{rev}  A.K.~Giem and K.S.~Novoselov, Nature Mat.~{\bf 6}, 183 (2007).

\bibitem{GPNJ}  V.~Chelanov and V.~Fal'ko, Phys.\ Rev.\ B {\bf 74}, 041403 (2006); L. M. Zhang and  M. M. Fogler, Phys.\ Rev.\ Lett.\ {\bf 100}, 116804 (2008).

\bibitem{GrExpMFP} E. H. Hwang, S. Adam, and S. Das Sarma, Phys. Rev. Lett. {\bf 98}, 186806 (2007); Y.-W. Tan, Y. Zhang, K.
Bolotin, Y. Zhao, S. Adam, E. H. Hwang, S. Das Sarma, H. L. Stormer, and P. Kim, Phys. Rev. Lett. {\bf 99}, 246803 (2007).
\end{thebibliography}
\end{document}